\documentclass[10.5pt,epsf]{article}
\usepackage{epsf}
\begin{document}
\title{ Inverted Hierarchical Model of Neutrino Masses Revisited}
\author{{\it{ Mahadev Patgiri}}$^{\dag}$
 {\it{and N. Nimai Singh}}$^{\ddag}$\\
$^{\dag}$ Department of Physics, Cotton College, Guwahati-781001, India\\
$^{\ddag}$ Department of Physics, Gauhati University, Guwahati-781014, India}
\date{}
\maketitle
\begin{abstract}
In this letter we highlight the inherent problems associated with  the inverted hierarchical model of neutrinos with only 
three generations and then suggest possible solutions within the MSSM.  
We discuss the new parametrization of the solar mixing angle which can identify the light side and dark side of the data. We  then argue whether  the inverted hierarchical neutrino mass matrix  can explain  the large mixing angle (LMA) MSW solution of the solar 
neutrino anomaly in the presence of an 
appropriate texture of charged lepton mass matrix.
In a model independent way we explore  such   specific form of the charged lepton mass 
matrix having a special 
structure in 1-2 block.  
  The contribution  to the solar mass splitting arising out of radiative corrections in MSSM,  is calculated, thus making the model stable under radiative corrections.
\end{abstract}

{\bf Introduction:} The data on solar neutrino 
oscillation has been traditionally shown on the $(\sin^{2}2\theta_{12},
\Delta m^{2}_{21})$ plane but   a  new parametrization on the 
  $(\tan^{2}\theta_{12},
\Delta m^{2}_{21})$ plane  with the usual  
sign convention $\Delta m^{2}_{21}=m^{2}_{2}-m^{2}_{1}>0$ has been recently suggested[1] to explain the solar data. In this new parametrization  the most recent result
from the SNO-Superkamiokande[2] favours the `` light side'', $\tan^{2}\theta_{12}<1$, and 
disfavours the ``dark side'', $\tan^{2}\theta_{12}>1$.
It  has also  imposed certain restrictions on the validity of the proposed 
neutrino mass models which are otherwise allowed in the old parametrization. 
 This can be understood from  the definition of 
leptonic mixing matrix $V_{MNS}$ where $\tan^{2}\theta_{12}=|V_{e2}|^{2}/|V_{e1}|^{2}<1$ 
implies $|V_{e2}|<|V_{e1}|$ for the usual sign convention  $\Delta m^{2}_{21}=m^{2}_{2}-m^{2}_{1} >0$, 
whereas such restriction does not exist in the definition of $\sin^{2}2\theta_{12}=4|V_{e1}|^{2}|V_{e2}|^{2}$. 

For our analysis in the present work, we  take here the  most general form of texture of the inverted neutrino masses, 
 which is assumed to be valid  at high scale
\begin{equation}
m_{LL}= 
\left(\begin{array}{ccc}
\delta & 1 & 1 \\
1 & \epsilon_{1} & \epsilon_{2} \\
1 & \epsilon_{2} & \epsilon_{1}
\end{array}\right) m_{0},
\end{equation}
Here we consider two possibilities. In case(i) we take  
$\delta = \epsilon_{1}=\epsilon_{2}=0$, leading to
 $\Delta m^{2}_{21}=0$. This represents  the case of perfect $L_{e}-L_{\mu}-L_{\tau}$ symmetry.  
 In case(ii) we have  $\delta \sim \epsilon_{1}\sim\epsilon_{2}\sim\lambda^{3}$, leading to $\Delta m^{2}_{21}>0$. 
 The quantum radiative corrections may modify the above two cases and  their low-energy  values are -
 case(i): $\Delta m^{2}_{sol}=\delta_{rad}$ when an appropriate charged lepton mass matrix is taken into consideration [3];  and case(ii): $\Delta m^{2}_{sol}=\Delta m^{2}_{21}+\delta_{rad}$.
 In MSSM the radiative correction term $\delta_{rad}$ is a negative quantity[4] and therefore, it will  lead
 to $\tan^{2}\theta_{12}>1$ in case (i). But in case (ii) if $\Delta m^{2}_{21}>\delta_{rad}$, 
the  radiative stability will still be maintained with $\tan^{2}\theta_{21}<1$. This point will be further examined in present work. 

{\bf Formalism and Results:}
 Expressing  $m_{LL}$ in the basis where the charged lepton mass matrix is  diagonal, we have    
$m^{\prime}_{LL}= V_{eL} m_{LL} V^{T}_{eL}, \ \ 
 m^{\prime  diag}_{LL}= V^{\prime}_{\nu L} m^{\prime}_{LL} V^{\prime T}_{\nu L},$ where 
$V_{MNS}= V^{\prime \dag}_{\nu L}$, and $V_{eL}$ is the diagonalizing matrix for charged lepton mass matrix.
The neutrino flavour eigenstate $\nu_{f}$ is related to the mass eigenstate $\nu_{i}$ by the relation 
$\nu_{f}=V_{fi}\nu_{i}$ 
and    the MNS mixing matrix is given by  $V_{fi}$ where  $f=\tau, \mu, e$ and   $i=1, 2, 3$.
 We take 
the usual convention of the neutrino mass eigenvalues $|m_{\nu1}|<|m_{\nu2}|$,  and this fixes 
the ordering of the first two columns of the leptonic mixing matrix $V_{MNS}$.
We present here 
 only one simple  case of the   left-handed Majorana neutrino mass matrix having $\delta=0,\epsilon_{2}=0,
\epsilon_{1}=\lambda^{3}$ for our demonstration in the present analysis,   
$m_{LL}$ with  $m_{0}=0.05eV$ in Eq.(1); and the neutrino mass eigenvalues $m_{i}= ( -1.4089, 1.4195,  0.01065)m_{0}$, 
$i= 1, 2, 3$, leading to the neutrino mass splittings 
$\Delta m^{2}_{21}=7.4952\times 10^{-5} eV^{2}$ and $\Delta m^{2}_{23}=5.0335\times 10^{-3} eV^{2}$. For our choice of the diagonal charged lepton mass matrix, the  MNS mixing matrix is  expressed as 
$$V_{MNS}=
\left(\begin{array}{ccc}
{ \frac{1}{\sqrt{2}}}+\alpha & {\frac{1}{\sqrt{2}}}-\alpha  & 0  \\
 {\frac{1}{2}}-\beta & -({\frac{1}{2}}+\beta) & -{\frac{1}{\sqrt{2}}} \\
  {\frac{1}{2}}-\beta & -({\frac{1}{2}}+\beta)  & {\frac{1}{\sqrt{2}}}
\end{array}\right)$$
where $\alpha=0.001335$ and $\beta=0.00094$. Here 
$\tan^{2}\theta_{12}=\frac{|{\frac{1}{\sqrt{2}}}-\alpha|^{2}}{|{\frac{1}{\sqrt{2}}}+\alpha|^{2}}<1$
for the usual convention  $\Delta m^{2}_{21}>0$. In case of  exact $L_{e}-L_{\mu}-L_{\tau}$ symmetry as in   case (i) where $\delta=\epsilon_{1,2}=0$ and 
   $\Delta m^{2}_{21}=0$, and  the above $V_{MNS}$  takes the form with $\alpha=0$ and $\beta=0$.  
In the  above example  the $V_{MNS}(=V^{\dag}_{\nu L})$ obtained from the  $m_{LL}$ alone 
fails to  explain the 
LMA MSW solution as it predicts the maximal solar mixings, and any small deviation in the texture of $m_{LL}$ will hardly 
affect the maximal value of $\tan^{2}\theta_{12}$. 
The last hope is that there could be a significant negative 
contribution to $\theta_{12}$ from $V_{eL}$ obtained from the diagonalisation of the charged 
lepton mass matrix $m_{l}$ having 
special structure in 1-2 block.  We wish to  examine here how
$\theta_{12}= (\theta^{\nu}_{12}- \theta^{e}_{12})$ can resolve the LMA MSW solar neutrino 
mixing scenario.

In a model independent way we parametrize the charged lepton mixing  $V_{eL}$ as 
\begin{equation}
V_{eL}=
\left(\begin{array}{ccc}
\bar{c}_{12} & \bar{s}_{12} & 0 \\ -\bar{s}_{12} & \bar{c}_{12} & 0 \\ 0 & 0 & 1
\end{array}\right)
\end{equation}
This gives a special form in the 1-2 block. We can reconstruct the symmetric  charged lepton mass matrix
using Eq.(2) from  the relation,
$m_{l}= V^{\dag}_{eL}m^{diag}_{l}V_{eR},$ 
where we consider $V_{eL}=V_{eR}$ for symmetric matrix.
The MNS mixing matrix $V_{MNS}=V_{eL}V_{\nu L}^{\dag}$ is now calculated as 
\begin{equation}
\left(\begin{array}{ccc}
({ \frac{1}{\sqrt{2}}}+\alpha)\bar{c}_{12}+ ({\frac{1}{2}}-\beta)\bar{s}_{12}  & ({ \frac{1}{\sqrt{2}}}-\alpha)\bar{c}_{12}- ({\frac{1}{2}}+\beta)\bar{s}_{12}  & -{\frac{\bar{s}_{12}}{\sqrt{2}}}  \\
 -({ \frac{1}{\sqrt{2}}}+\alpha)\bar{s}_{12}+ ({\frac{1}{2}}-\beta)\bar{c}_{12}    &  -({ \frac{1}{\sqrt{2}}}-\alpha)\bar{s}_{12}- ({\frac{1}{2}}+\beta)\bar{c}_{12}  & -{\frac{\bar{c}_{12}}{\sqrt{2}}}  \\ 
  {\frac{1}{2}}-\beta & -({\frac{1}{2}}+\beta)  & {\frac{1}{\sqrt{2}}}
\end{array}\right)
\end{equation}
Now we obtain the following parameters 
$$\tan^{2}\theta_{21}=\frac{|({\frac{1}{\sqrt{2}}}\bar{c}_{12}-\frac{1}{2}\bar{s}_{12})-\alpha(\bar{c}_{12}+\bar{s}_{12})|^{2}}{| ({\frac{1}{\sqrt{2}}}\bar{c}_{12}+\frac{1}{2}\bar{s}_{12})+\alpha(\bar{c}_{12}-\bar{s}_{12})|^{2}},$$
$$\sin^{2}2\theta_{23}=\bar{c}_{12}^{2},\ \ \
|V_{e3}|=|\frac{\bar{s}_{12}}{\sqrt{2}}|,\ \ \
|m_{ee}|\simeq |m_{1}|\sqrt{2}\bar{s}_{12}\bar{c}_{12}.$$

If ${\Delta_{23} =1-\sin^{2}2\theta_{23}}$, is the small deviation from the completely maximal value of the atmospheric neutrino mixings, 
then the above relations take the following new forms:

\begin{equation}
\tan^{2}\theta_{21} = |1-\sqrt{8\Delta_{23}} + 3\Delta_{23}|
\end{equation}

\begin{equation}
|V_{e3}|=\sqrt{\frac{\Delta_{23}}{2}},  |m_{ee}|=|m_{1}|\sqrt{2\Delta_{23}}
\end{equation}

Where ${\alpha}$ in Eq.(4)has been suppressed as it contributes insignificantly to the solar mixing.

Table-1: Few reprsentative cases for the values of the quantities in Eqs.(4) and (5) obtained 
for the different values of ${\Delta_{23}}$ as input.
\begin{center}
\begin{tabular}{cccccc}\hline
Case &  ${\Delta_{23}}$  & ${\sin^{2}2\theta_{23}}$ & ${|V_{e3}|}$ & ${\tan^{2}\theta_{21}}$ & ${|m_{ee}| eV}$\\
\hline 
I &  .02  &  0.98      &  0.10 & 0.66 & 0.014\\
II & .04 & 0.96  & 0.14 & 0.55 & 0.020\\
III& .05  & 0.95 & 0.16 & 0.52 & 0.022\\
IV & .06 & 0.94 & 0.17  & 0.49 & 0.024\\
\hline
\end{tabular}
\end{center}
The Table-1 shows that in Cases II and III, solar mixings together with  solar mass splitting
${\Delta m_{21}^{2} =7.49\times 10^{-5}}$ ${eV^{2}}$ are in excellent agreement with the best fit ${\tan^{2}\theta_{21}=0.56}$ obtained in KamLand experiment[8].
The values of ${|V_{e3}|}$ are also in accord with the CHOOZ experiment. 

Finally we check the stability of the neutrino mass matrix under RGEs. Following the standard procedure[4,5] we express $m_{LL}$ in terms 
of $K$, the coefficient of the dimension 
five neutrino mass operator  in a scale-dependent manner
$m_{LL}(t)= v^{2}_{u}(t) K(t)$
where $t=ln(\mu)$ and $v_{u}(t)$ is the scale-dependent[5] vacuum expectation value (VEV) 
$v_{u}= v_{0} \sin\beta$, $v_{0}= 174 GeV$. In the basis where the charged lepton mass matrix is made  diagonal,
we can write the above expression as
$m^{\prime}_{LL}(t)= v^{2}_{u}(t) K^{\prime}(t)$
where $K^{\prime}(t)=V_{eL}K(t)V_{eL}^{T}$ is the coefficient of the dimension five neutrino mass operators in the basis 
where the charged lepton mass matrix is diagonal. The approximate analytical solution of the relevant RGEs in MSSM, is given by the  following form 
\begin{equation}
m^{\prime}_{LL}(t_{0})=R.diag(1,1,\alpha).m^{\prime}_{LL}(t_{u}).diag(1,1,\alpha)
\end{equation}
We have the expressions  $R= e^{\frac{9}{10}I_{g_{1}}(t_{0})} e^{\frac{9}{2}I_{g_{2}}(t_{0})}\simeq 1$, \ \ \
$\alpha=e^{-t_{\tau}(t_{0})}=1-r$ where the approximate solution is [4,5,6] 
$$r=1-(m_{t}/M_{u})^{(1+\tan^{2}\beta)(m_{\tau}/2\sqrt{2}\pi v)^{2}}$$
with $v=245.4GeV$. The low-energy solar mass splitting $\Delta m^{2}_{sol}$ can be expressed in terms of high energy solar mass splitting $\Delta m^{2}_{21}$ and the radiative correction effect $\delta_{rad}$ as 
\begin{equation}
\Delta m^{2}_{sol}= \Delta m^{2}_{21}+ \delta_{rad},\\
\delta_{rad}= -4 r |m_{1}|^{2}\sin^{2}2\theta_{23} \cos 2\theta_{21}.
\end{equation}
In MSSM the quantity  $r$ is positive and thus we have $\Delta m^{2}_{sol}<\Delta m^{2}_{21}$. For example, We obtain the numerical values $\delta_{rad}=-0.96\times 10^{-5}eV^{2}$ corresponding to the CaseIII in Table-1. The  mass ratio $|m_{2}|/|m_{1}|$  also decreases with energy scale. The model is stable under radiative corrections.

\end{document}